\documentclass[aps,prb,amsmath,twocolumn]{revtex4-2}
\usepackage{graphicx}
\usepackage{bm}

\begin{document}
\title{Josephson dynamics at high transmissions: Perturbation theory}
\author{Artem V. Galaktionov}
\affiliation{I.E. Tamm Department of Theoretical Physics, P.N. Lebedev Physical Institute, 119991 Moscow, Russia}
\author{Andrei D. Zaikin}
\affiliation{I.E. Tamm Department of Theoretical Physics, P.N. Lebedev Physical Institute, 119991 Moscow, Russia}
\affiliation{National Research University Higher School of Economics, 101000 Moscow, Russia}

\date{\today}
\begin{abstract}
We theoretically analyze Josephson dynamics of superconducting weak links with transmissions ${\mathcal T}$ not much smaller than unity at subgap bias voltages $V$. Employing the effective action approach combined with the Keldysh technique we develop a regular perturbation theory in ${\mathcal R}=1-{\mathcal T}$ and derive the first order correction to the current across the weak link which consists of two different contributions.  One of them is negative effectively corresponding to a decrease of the excess current at small $V$ due to breaking of the multiple Andreev reflection cycle by normal reflection for some subgap quasiparticles. These quasiparticles, in turn, generate the second -- Josephson-like -- contribution to the current which increases with decreasing $V$ down to very small voltages where the perturbation theory in ${\mathcal R}$ ceases to be valid. Some of the above features are not reproduced within the physical picture involving Landau-Zener tunneling between subgap Andreev states.
\end{abstract}
\maketitle

\section{Introduction}

Josephson dynamics beyond the tunneling limit involves a non-trivial interplay between superconductivity and non-equilibrium effects which can in general be described only with the aid of complicated many-body techniques possibly combined with numerical methods. In some special cases one can also proceed analytically. An example of such situation has to do with ac Josephson effect in short ballistic superconductor-normal metal-superconductor (SNS) junctions or superconducting quantum point contacts at full transmissions  \cite{Uwe,ab1}. In this case
for bias voltages $V$ well below the superconducting gap of the electrodes, $eV \ll \Delta$, one recovers the following expressions for the time-dependent current across the system
\begin{equation}
I(t)\equiv I_0(t)=I_c \left| \sin eVt\right|\,{\rm sgn}\,V
\label{1}
\end{equation}
and for the $I-V$ curve
\begin{equation}
\overline{I_0}=\frac{2I_c}{\pi} \,{\rm sgn}\,V,
\label{2}
\end{equation}
where
\begin{equation}
I_c=e\Delta \tanh\left(\frac{\Delta}{2T}\right)\,
\label{3}
\end{equation}
is the dc Josephson critical current  \cite{KO} of the structure at temperature $T$. Note that for simplicity in Eqs. (\ref{1}), (\ref{2}) we neglected the growing with $V$ term that is small in the parameter $eV/\Delta$. Equations (\ref{1})-(\ref{3}) hold for a superconducting point contact with a single transport channel at full transmission ${\mathcal T}=1$ and are trivially generalized to the case of an arbitrary number of channels.
Note that here and below $-e$ is the electron charge, and we set the Planck and Boltzmann constants equal to unity ($\hbar=k_B=1$).

The charge transfer in such superconducting contacts is essentially governed by the mechanism of multiple Andreev reflection (MAR) \cite{MAR}
which yields, e.g., the large excess current on the $I-V$ curve (\ref{2}) already at small voltages as well as a non-trivial current-phase relation (\ref{1}). Remarkably, it was demonstrated \cite{ab1,ab2} that the same results can also be recovered without involving the physical picture of MAR but operating only with occupation probabilities of subgap Andreev bound states $\pm \epsilon_A$ (see Fig. 1). The corresponding kinetic equation that controls dynamics of these  occupation probabilities is supplemented with the boundary conditions \cite{ab1,ab2} for the superconducting phase difference $\chi$ across the junction resulting from discrete Andreev levels merging with the continuum at $\chi =0, \pm 2\pi, ...$

\begin{figure}
\includegraphics[width=8cm]{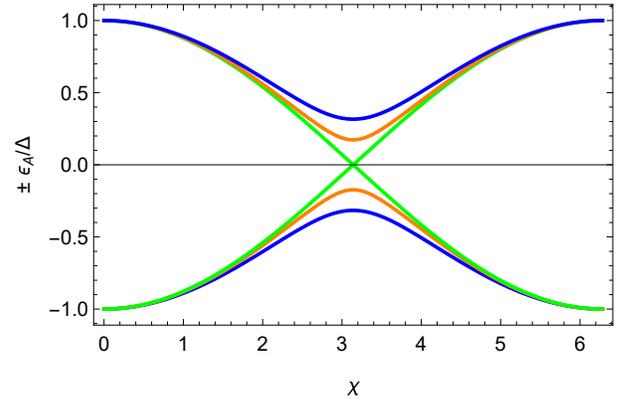}
\caption{A pair of Andreev levels $\pm \epsilon_A(\chi )$ in a superconducting quantum point contact with ${\mathcal T}=1, 0.97$ and 0.9 shown respectively by green, orange and blue lines.}
\end{figure}

The description of Josephson dynamics at subgap voltages employing the effective basis of Andreev levels was also extended \cite{ab1,ab2} to the case of non-zero reflection coefficients ${\mathcal R}\equiv 1-{\mathcal T}>0$, which otherwise was treated with the aid of numerics \cite{Chalmers,Madrid}.  For non-zero ${\mathcal R}$ the gap between two Andreev levels develops (see Fig. 1), and it was suggested in Refs. \onlinecite{ab1,ab2} to include the effect of  Landau-Zener tunneling between these levels with the probability $p=\exp (-\pi {\mathcal R}\Delta/e|V|)$  as an extra boundary condition  at $\chi=\pi\mod(2\pi)$. Following this scenario one concludes that, since for $e|V| \ll  \pi {\mathcal R}\Delta$ the probability of Landau-Zener tunneling is exponentially small $p \ll 1$, the system should remain in the lowest Andreev state and, hence, the current $I(t)$ is essentially described by a time-dependent version of the Kulik-Omelyanchuk current-phase relation \cite{KO}. On the other hand, in the opposite limit $e|V| \gg  \pi {\mathcal R}\Delta$ (though $e|V| \ll 2\Delta$ also implying ${\mathcal R}\ll 1$) the probability $p$ becomes close to unity, both Andreev levels get occupied and the current $I(t)$ can be evaluated perturbatively in $1-p$. In this limit Eq. (11) of Ref. \onlinecite{ab1} yields
\begin{equation}
I(t)=I_0(t)-\delta I(t),  \quad \delta I(t)=\pi \overline{\delta I} F(eVt),
\label{cab1}
\end{equation}
where
\begin{equation}
\overline{\delta I}\simeq \frac{2 {\mathcal R} \Delta}{eV}I_c
\label{corr}
\end{equation}
denotes the correction to the average current (i.e. $\overline{I}=\overline{I_0}-\overline{\delta I}$) and
\begin{equation}
F(x)= F_{LZ}(x)=\Theta\left(x- \frac{\pi}{2}\right)\sin x, \quad 0\le x\le \pi
\label{cab2}
\end{equation}
is the $\pi$-periodic function that accounts for Landau-Zener tunneling and $\Theta(x)$ is the Heaviside step function.  Note that the current correction $\delta I(t)$ defined in Eqs. (\ref{cab1}), (\ref{cab2}) becomes discontinuous at $2eVt=\pi\mod(2\pi)$.

The above reduced description of adiabatic Josephson dynamics operating only with subgap Andreev states was later employed by many authors in different physical contexts, see, e.g., Refs. \onlinecite{Yakovenko,Chalmers2,FK,SSt} as well as many other publications. While this description is quite simple and intuitively appealing, the question remains if it is fully equivalent to the standard physical picture of the charge transfer in terms of MAR. This question will be addressed in our present work.

Note that the applicability condition of the perturbation theory in ${\mathcal R} \ll 1$ can easily be reconstructed also within the framework of the physical picture of MAR. Indeed, in order for the current $I(t)$ to be only weakly disturbed as compared to $I_0(t)$ (\ref{1}) the total normal reflection probability  ${\mathcal R}n$ within the full MAR cycle of $n$ traverses across the weak link (see Fig. 2) should remain much smaller than unity. Since $n \sim 2\Delta /e|V|$, we immediately recover the condition $e|V| \gg 2{\mathcal R}\Delta$ essentially equivalent to that derived within the Landau-Zener tunneling scenario.

\begin{figure}
\includegraphics[width=8cm]{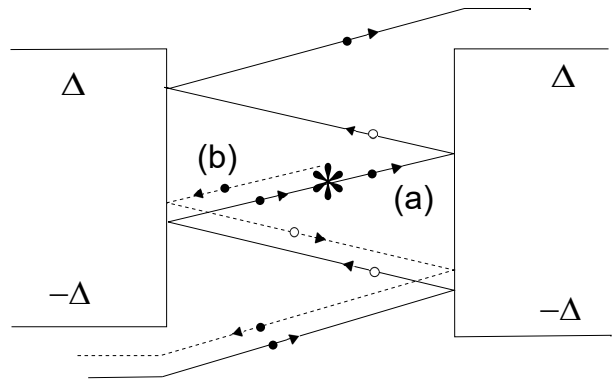}
\caption{A typical MAR trajectory in a highly transparent superconducting weak link biased by a small constant voltage $V \ll \Delta/e$. A quasiparticle (hole) suffers successive Andreev reflections at both NS interfaces (solid line) until it gets normally scattered at the point marked by a star. After this scattering event the quasiparticle either (a) passes the barrier with the probability ${\mathcal T}=1-{\mathcal R}$  and completes the MAR cycle (solid line) or (b) gets out of the MAR cycle with the probability ${\mathcal R}$  and further propagates along the time-reversed trajectory indicated by the dashed line. Since normal scattering occurs for each of $n$ traverses across the weak link the total probability for the quasiparticle to leave the full MAR cycle is $1-{\mathcal T}^n$ which equals to ${\mathcal R}n$ as long as ${\mathcal R}n \ll 1$. }
\end{figure}

Under this condition below we will develop a regular perturbation theory in ${\mathcal R}$ and evaluate the first order correction to the current
$\delta I(t)$ both analytically and numerically. We demonstrate that the function $\delta I(t)$ is continuous at any $t$. Expressing it in terms of Fourier series, we have
\begin{equation}
\delta I(t)=\pi \overline{\delta I}\sum_{l=-\infty}^\infty F_l e^{-2ileVt}, \quad F_{-l}=F_l^*.\label{seq}
\end{equation}
Within the framework of our analysis we arrive at the result for the average current  $\overline{\delta I}$ consistent with Eq. (\ref{corr}).
At the same time, all Fourier coefficients $F_l$ with $|l|> 1$ exhibit qualitatively different behavior from that found for the analogous Fourier coefficients of the function $F_{LZ}(x)$ (\ref{cab2}).

Our paper is organized as follows. In Section II we develop a regular perturbation theory that allows to evaluate the first order in ${\mathcal R}$ correction $\delta I(t)$ to the current flowing across a superconducting weak link. Section III is devoted to the calculation of $\delta I(t)$ in the limit of small constant voltage bias $|V| \ll 2\Delta/e$. The results are further discussed in Section IV.  Additional information employed in our  numerical calculation is presented in Appendix A.

\section{Perturbation theory}

In our subsequent analysis, we will closely follow the procedure outlined in detail in Ref. \onlinecite{Sha}, i.e. we will employ the combination of the Keldysh technique and the effective action approach. In order to describe charge transfer across our superconducting weak link with arbitrary distribution of transmission probabilities ${\mathcal T}_k\equiv 1-{\mathcal R}_k$ over its conducting channels $k$ we routinely use the general expression for the effective action \cite{Z94,SN} that can be written in the form
\begin{eqnarray}
&& iS_t[\varphi]=\frac{1}{2}\sum_k {\rm Tr}\,\ln\left[ 1+\frac{{\mathcal T}_k}{4}\left(\left\{\check Q_L(\varphi),\check Q_R\right\}-2 \right)\right]\nonumber\\ &&=\frac{1}{2}\sum_k {\rm Tr}\,\ln\left[\frac{1-{\mathcal R}_k}{4}\left(\check Q_L+\check Q_R\right)^2+{\mathcal R}_k \right].\label{ea}
\end{eqnarray}
The summation over all conducting channels $k$ is implied in Eq. (\ref{ea}), and $\{...,...\}$ stands for the anticommutator. The product of (defined in Keldysh and Nambu spaces) $4\times 4$ matrices $\check Q_L$ and $\check Q_R$, describing respectively the left and the right superconducting reservoirs implies the time convolution:
\begin{equation}
\left( \check Q_L\circ \check Q_R\right) (t',t'')=\int\limits_{-\infty}^{\infty} dt \check Q_L(t',t) \check Q_R(t,t''),
\end{equation}
and the matrices $\check Q_{L,R}$  obey the standard normalization condition $\left( \check Q_L\circ \check Q_L\right) (t',t'')=\delta(t'-t'')$, where $\delta(t)$ is the Dirac delta function. This normalization condition was directly employed in order to cast the action to the form presented the second line in Eq. (\ref{ea}) which serves as a convenient starting point of our perturbation theory.

Assuming that all reflection coefficients ${\mathcal R}_k$ are much smaller than unity we may formally limit our analysis to the first order in ${\mathcal R}_k$ and further rewrite Eq.(\ref{ea}) as
\begin{eqnarray}
&& iS_t[\varphi]=\frac{1}{2}\sum_k {\rm Tr}\,\ln \left[\frac{1}{4}\bigg(\left(1-\frac{{\mathcal R}_k}{2} \right)\left(\check Q_L+\check Q_R\right)\right.\nonumber\\&&  \left.+2{\mathcal R}_k \left(\check Q_L+\check Q_R\right.)^{-1}\bigg)^2 \right].
\end{eqnarray}
Introducing the matrix
\begin{equation}
\check Q=\frac{1}{2}\check I\left(\check Q_L+\check Q_R\right), \quad \check I =\left(\begin{array}{cc} \hat\tau_3& 0\\ 0& -\hat\tau_3\end{array} \right),
\end{equation}
where $\hat\tau_3$ stands for the Pauli matrix, we can convert the action to the form
\begin{equation}
iS_t[\varphi]=\sum_k {\rm Tr}\,\ln \left[\left(1-\frac{{\mathcal R}_k}{2} \right)\check Q+\frac{{\mathcal R}_k}{2}\check I\check Q^{-1}\check I \right].
\end{equation}
It is straightforward to check that at $t\rightarrow t'$ one has $\check Q(t,t')\approx \delta(t-t')$, i.e. the expansion of the logarithm is properly organized.

Following \cite{Sha} let us decompose the matrix $\check Q$ as
\begin{equation}
\check Q=\check Q_0+\check Q_1.
\end{equation}
The matrix $\check Q_0$ reads
\begin{equation}
\check Q_0=\left( \begin{array}{cc} \hat a^R & \hat a^K \\ 0 & -\hat a^A \end{array}\right), \label{q0s}
\end{equation}
where
\begin{widetext}
\begin{equation}
\hat a^{R,A,K}(t,t')=\left(\begin{array}{cc} g^{R,A,K}(t,t')\cos\left[\frac{\varphi_+(t)-\varphi_+(t')}{2} \right] & f^{R,A,K} (t,t')\cos\left[\frac{\varphi_+(t)+\varphi_+(t')}{2} \right] \\ f^{R,A,K}(t,t')\cos\left[\frac{\varphi_+(t)+\varphi_+(t')}{2} \right]& g^{R,A,K}(t,t')\cos\left[\frac{\varphi_+(t)-\varphi_+(t')}{2} \right]\end{array}\right), \label{adef}
\end{equation}
\end{widetext}
$g^{R,A,K}$ and $f^{R,A,K}$ denote respectively normal and anomalous retarded (R), advanced (A) and Keldysh (K) quasiclassical Green functions of a superconductor \cite{ZRev} and -- in the absence of electron-electron interactions -- $\varphi_+(t)=\varphi_0+e\int_0^t V(t')dt'$ is a half of the time-dependent superconducting phase difference across our weak link.

The matrix $\check Q_1$ has the form
\begin{eqnarray}
&& \check Q_1(t,t')=\frac{\varphi_-(t)}{4}\check B(t,t')+\\ && \left(\begin{array}{cc} 0& \hat \tau_3\\ \hat \tau_3& 0 \end{array} \right)\check B(t,t')\left(\begin{array}{cc} 0& \hat \tau_3\\ \hat \tau_3& 0 \end{array} \right)\frac{\varphi_-(t')}{4},\nonumber
\end{eqnarray}
where
\begin{equation}
\check B(t,t')=\left(\begin{array}{cc} 0 & -\hat b^A(t,t')\\ \hat b^R(t,t') & \hat b^K(t,t') \end{array} \right),\label{bsdef}
\end{equation}
$\varphi_-$ is the "quantum" part of the phase difference and
\begin{widetext}
\begin{equation}
\hat b^{R,A,K}(t,t')=\left(\begin{array}{cc} g^{R,A,K}(t,t')\sin\left[\frac{\varphi_+(t)-\varphi_+(t')}{2} \right] & f^{R,A,K}(t,t')\sin\left[\frac{\varphi_+(t)+\varphi_+(t')}{2} \right] \\ f^{R,A,K}(t,t')\sin\left[\frac{\varphi_+(t)+\varphi_+(t')}{2} \right] & g^{R,A,K}(t,t')\sin\left[\frac{\varphi_+(t)-\varphi_+(t')}{2} \right]\end{array} \right).\label{bexp}
\end{equation}
\end{widetext}

Neglecting electron-electron interactions we can define the current across our weak link by means of the formula
\begin{equation}
I(t)= -e\frac{\delta S_t}{\delta\varphi_-(t)}|_{\varphi_-(t)=0}.
\end{equation}
Combining this formula with the above expressions for the action $S_t$, in the first order in ${\mathcal R}_k$ we get
\begin{eqnarray}
&& I(t)=\frac{ie}{4}\sum_k {\rm Tr}\left[\check B \check Q_0^{-1}\left( 1-{\mathcal R}_k\left(\check I\check Q_0^{-1} \right)^2\right) +\right.\nonumber\\&& \left. \left( 1-{\mathcal R}_k\left(\check Q_0^{-1}\check I \right)^2\right)\check Q_0^{-1}\check B^\prime \right](t,t),\label{gc}
\end{eqnarray}
where we also defined
\begin{equation}
\check B^\prime(t,t')=\left(\begin{array}{cc} 0& \hat \tau_3\\ \hat \tau_3& 0 \end{array} \right)\check B(t,t')\left(\begin{array}{cc} 0& \hat \tau_3\\ \hat \tau_3& 0 \end{array} \right). \label{bpsdef}
\end{equation}

\section{Constant voltage limit}

While the above perturbative expression for the current (\ref{gc}) generally holds for an arbitrary dependence of the applied voltage $V(t)$ on time, below we will specifically restrict our analysis to the time-independent voltage bias limit $V(t)\equiv V$.  In this special case the solution for the inverse matrix $\check Q_0^{-1}$ has already been constructed earlier \cite{Sha}. It is convenient to write the components of this matrix using the following expansion in terms of the voltage harmonics
\begin{equation}
x(t,t')=\sum_{n=-\infty}^\infty \int\frac{d\epsilon}{2\pi} x(\epsilon,n) e^{-i\epsilon(t-t')} e^{-i ne V(t+t')/2}.\label{feq}
\end{equation}
The retarded component of the inverse matrix $\check Q_0^{-1}$ has been demonstrated to acquire the structure \cite{Sha}
\begin{equation}
\left(\begin{array}{cc}\zeta^R(\epsilon,2n)&  \zeta^R(\epsilon,2n+1)\\ \zeta^R(\epsilon,2n+1)& \zeta^R(\epsilon,2n)\end{array}  \right),\label{zets}
\end{equation}
i.e. the components with even index are diagonal and those with odd index are off-diagonal. We also introduce components with shifted energy arguments. Here and below for brevity we denote these components by tilde, i.e.  $\tilde\zeta^R (\epsilon,n)= \zeta^R \left(\epsilon+(neV/2),n\right)$.  We have
\begin{equation}
\tilde\zeta^R\left(\epsilon+\frac{eV}{2},l \right)=\left\{ \begin{array}{l} (-1)^l\prod_{1\le k \le l}a^R(\epsilon+eVk),\; {\rm if}\; l>0, \\ 1,\quad {\rm if}\; l=0, \\ (-1)^l\prod_{l+1\le k \le 0}a^R(\epsilon+eVk),\; l<0, \end{array}\right.\label{iel}
\end{equation}
where
\begin{equation}
a^R(\epsilon)=\frac{f^R(\epsilon)}{1+ g^R(\epsilon) }\label{are}
\end{equation}
defines the Andreev reflection amplitude. This combination is also involved in the standard Riccati parametrization for the Green functions \cite{ZRev}
\begin{equation}
f^R=\frac{2 a^R}{1-(a^R)^2},\quad g^R=\frac{1+(a^R)^2}{1-(a^R)^2}. \label{rp}
\end{equation}

Having in mind that
\begin{equation}
g^{R,A}(\epsilon)=\frac{\epsilon\pm i\theta}{\xi^{R,A}(\epsilon)},\quad f^{R,A}(\epsilon)=\frac{\Delta}{\xi^{R,A}(\epsilon)},\label{arch}
\end{equation}
where $\xi^{R,A}(\epsilon)=\pm\sqrt{(\epsilon\pm i\theta)^2-\Delta^2}$ and $\theta$ phenomenologically controls the strength of inelastic relaxation, from
Eq. (\ref{are}) in the limit $\theta \to 0$ we obtain
\begin{eqnarray}
a^R(\epsilon)=\frac{\epsilon}{\Delta}-\frac{i\sqrt{\Delta^2-\epsilon^2}}{\Delta}=\exp\left( -i\arccos \frac{\epsilon}{\Delta}\right)
\label{arc}
\end{eqnarray}
for $|\epsilon|<\Delta$ and
\begin{equation}
a^R(\epsilon)=\frac{{\rm sgn}\,\epsilon}{\Delta}\left( |\epsilon|-\sqrt{\epsilon^2-\Delta^2}\right)
\label{leps}
\end{equation}
for $|\epsilon|>\Delta$.

Note that the multiplicative structure (\ref{iel}) is associated with the process of MAR. Formally it results from the multiplicative structure of the inverse symmetric tridiagonal matrix, as it is discussed in Ref. \onlinecite{M}. The advanced and Keldysh components of the inverse matrix have also been established and analyzed in Ref. \onlinecite{Sha}. Here, however, it is sufficient for our purposes to restrict our attention to the retarded matrix component.

Making use of Eq.(\ref{gc}), we obtain
\begin{eqnarray}
&& \delta I(t)=\frac{ie}{2}{\mathcal R}\left(b^R_gY^K_g+b^R_fY^K_f+b^K_gY^A_g + b^K_fY^A_f\right.\nonumber\\ && \left. +Y^R_gb^K_g-Y^R_fb^K_f-Y^K_gb^A_g+ Y^K_fb^A_f \right)_{t,t}.\label{exp1}
\end{eqnarray}
Note that from here on for simplicity we only consider the case of a weak link with a single conducting channel characterized by the reflection coefficient ${\mathcal R} \ll 1$. Generalization to an arbitrary number of channels simply amounts to substitute ${\mathcal R} \to \sum_k {\mathcal R}_k$ in any of our final results.

The symmetric $2\times 2$ matrices in Eq. (\ref{bexp}) can be expressed in the form
\begin{equation}
\hat b^R=b^R_g\hat 1+b^R_f\hat\tau_2,
\end{equation}
with
\begin{eqnarray}
&& b^R_g(t,t')= g^{R}(t,t')\sin\left[\frac{\varphi_+(t)-\varphi_+(t')}{2} \right],\\ && b^R_f=f^{R}(t,t')\sin\left[\frac{\varphi_+(t)+\varphi_+(t')}{2} \right],\nonumber
\end{eqnarray}
while the matrix $\check Y$ involved in Eq. (\ref{exp1}) is defined as
\begin{equation}
\check Y=\check Q_0^{-1}\check I \check Q_0^{-1}\check I \check Q_0^{-1}.
\end{equation}

Let us introduce the notations
\begin{eqnarray}
&& \check Q_0^{-1}=\left(\begin{array}{cc}\hat X^R & \hat X^K \\ 0 & \hat X^A \end{array}\right),\; \hat X^R=\left( \hat a^R\right)^{-1},\label{mgdef}\\
&&  \hat X^A=-\left( \hat a^A\right)^{-1},\; \hat X^K=-\hat X^R\circ \hat a^K\circ \hat X^A. \nonumber
\end{eqnarray}
The components of the matrix $X^R$ are already specified in Eqs. (\ref{zets}), (\ref{iel}), whereas for the components of the  matrix $\check Y$ we have
\begin{eqnarray}
&& Y^R_g=X^R_gX^R_gX^R_g+X^R_fX^R_gX^R_f-X^R_f X^R_fX^R_g\nonumber\\&& -X^R_gX^R_fX^R_f,\label{x1}\\
&& Y^R_f=X^R_gX^R_gX^R_f+X^R_fX^R_gX^R_g-X^R_f X^R_fX^R_f\nonumber\\&&  -X^R_gX^R_fX^R_g,\label{x11}\\
&& Y^A_g=X^A_gX^A_gX^A_g+X^A_fX^A_gX^A_f-X^A_f X^A_fX^A_g\nonumber\\ && -X^A_gX^A_fX^A_f,\label{x111}\\
&& Y^A_f=X^A_gX^A_gX^A_f+X^A_fX^A_gX^A_g-X^A_f X^A_fX^A_f\nonumber\\ && -X^A_gX^A_fX^A_g.\label{x1111}
\end{eqnarray}
The expressions for $Y^K$-components turn out to be somewhat lengthier. They are
\begin{eqnarray}
&& Y^K_g=X^R_gX^R_gX^K_g+X^R_fX^R_gX^K_f-X^R_f X^R_fX^K_g\nonumber\\&& -X^R_g X^R_fX^K_f
+X^K_gX^A_gX^A_g+X^K_fX^A_gX^A_f\nonumber \\&&-X^K_f X^A_fX^A_g-X^K_gX^A_fX^A_f
+X^R_gX^K_fX^A_f\nonumber\\ && +X^R_fX^K_fX^A_g-X^R_g X^K_gX^A_g-X^R_fX^K_gX^A_f\label{x2}
\end{eqnarray}
and
\begin{eqnarray}
&& Y^K_f=X^R_gX^R_gX^K_f+X^R_fX^R_gX^K_g-X^R_f X^R_fX^K_f\nonumber\\&&-X^R_gX^R_fX^K_g +X^K_gX^A_gX^A_f+X^K_fX^A_gX^A_g\nonumber \\ &&-
X^K_fX^A_fX^A_f -X^K_gX^A_fX^A_g  +X^R_gX^K_fX^A_g \nonumber\\ &&+X^R_fX^K_fX^A_f-X^R_g X^K_gX^A_f-X^R_fX^K_gX^A_g.\label{x3}
\end{eqnarray}

Consider the energies $-\Delta-eV/2<\epsilon<-\Delta+eV/2$ which merely contribute to the current harmonics.
One can use the following property of the convolution $z=fg$ with energy-shifted components,
\begin{equation}
\tilde z(\epsilon,n)=\sum_{k+l=n}\tilde f(\epsilon+leV,k)\tilde g(\epsilon,l).\label{cru2}
\end{equation}
Introducing the combinations $T_1^R=X^R_gX^R_g-X^R_fX^R_f$ and $T_2^R=X^R_gX^R_f-X^R_fX^R_g$ and limiting the summation in Eq.(\ref{cru2}) to $k \ge 0, l\ge 0$ one can demonstrate that the following identities hold for $n>0$
\begin{equation}
\tilde T_1(\epsilon,n)=\tilde X^R_g(\epsilon,n),\quad \tilde T_2(\epsilon,n)=0.\label{cap}
\end{equation}
While deriving Eq. (\ref{cap}) we chose to neglect Andreev reflection at overgap energies, which effectively amounts to set $a^R(\epsilon)=0$ at $|\epsilon|>\Delta$.
 This approximation was used, e.g., in Refs. \onlinecite{Uwe,ab1} based on the fact that deep in the interesting for us subgap voltage regime $eV \ll \Delta$ higher order MAR processes (and, hence, higher powers of $a^R(\epsilon)$) determine the current across the contact. This approximation just simplifies the calculation and -- as will be demonstrated below -- is by no means crucial for our results and conclusions.

Employing the relations (\ref{cap}), we obtain
\begin{equation}
\tilde Y^R_g(\epsilon,2n)=(1+|n|)\tilde\zeta^R(\epsilon,2n).\label{yaa}
\end{equation}
and
\begin{eqnarray}
&& \tilde Y^R_f(\epsilon,2n+1)=(1+n)\tilde \zeta^R(\epsilon,2n+1),\nonumber\\
&& \tilde Y^R_f(\epsilon,-2n-1)=(1+n)\tilde \zeta^R(\epsilon,-2n-1)\label{ya}
\end{eqnarray}
for $n\ge 0$. Let us emphasize the presence of an additional factor $n$ in Eqs. (\ref{yaa}), (\ref{ya}), which becomes particularly important in the limit of small voltages $eV \ll \Delta$.

More generally, at sufficiently large $n$ one may write
$\tilde Y^R_{g,f}(\epsilon,n)\approx \alpha (|n|/2)\tilde \zeta^R(\epsilon,n)$, where the dimensionless prefactor $\alpha$ can be determined explicitly only if one abandons the approximation $a^R(\epsilon)=0$ at $|\epsilon|>\Delta$ employed above. We will not proceed with this (much more complicated) analysis here and simply determine the prefactor $\alpha$ numerically (see below).

The Keldysh component of the matrix $\check Y$ can be written as
\begin{eqnarray}
&&\hat Y^K=-\hat X^R\hat\tau_3\hat X^R\hat\tau_3\hat X^R \hat a^K\hat X^A\label{systr}\\&& - \hat X^R \hat a^K\hat X^A\hat\tau_3\hat X^A\hat\tau_3\hat X^A+ \hat X^R\hat\tau_3\hat X^R \hat a^K\hat X^A\hat\tau_3\hat X^A.\nonumber
\end{eqnarray}
The first two summands can be expressed in terms of $\hat Y^R$ and $\hat Y^A$, while the third summand can be neglected in the low voltage limit as it does not contain an additional $n$-factor corresponding to higher order MAR processes.

Combining all the above expressions, after some algebra we obtain the linear in ${\mathcal R}$ correction to the current in the form (\ref{seq}),
where
\begin{equation}
\overline{\delta I}=\frac{2 \alpha {\mathcal R} \Delta^2}{\pi V}\tanh\left( \frac{\Delta}{2T}\right),
\label{seq1}
\end{equation}
and
\begin{equation}
F_l=\frac{1}{2\pi}\int\limits_{-1}^1 dx (1-x)\exp(2il\arccos x).
\label{seq2}
\end{equation}

Equations (\ref{seq1}) and (\ref{seq2}) combined with Eq. (\ref{seq}) represent the central result of our present work. They determine the correction to the current $I(t)$ across weakly reflecting superconducting point contacts in the interesting for us voltage range  ${\mathcal R} \Delta \ll eV \ll 2\Delta$.

Evaluating the integral in Eq. (\ref{seq2}), we obtain
\begin{equation}
{\rm Re}\,F_l=\frac{1}{\pi(1-4l^2)},\quad {\rm Im}\,F_l=\left\{\begin{array}{c} \mp\frac{1}{8},\quad{\rm if}\;l=\pm 1\\ 0 ,\quad{\rm if}\;l\neq\pm 1\end{array}\right.
\end{equation}
and recover the $\pi$-periodic function $F(eVt)$ in the form
\begin{equation}
F(x)=\frac{1}{2}\left| \sin x\right| -\frac{1}{4}\sin(2x), \quad 0\le x\le \pi . \label{tf}
\end{equation}

\begin{figure}
\includegraphics[width=8cm]{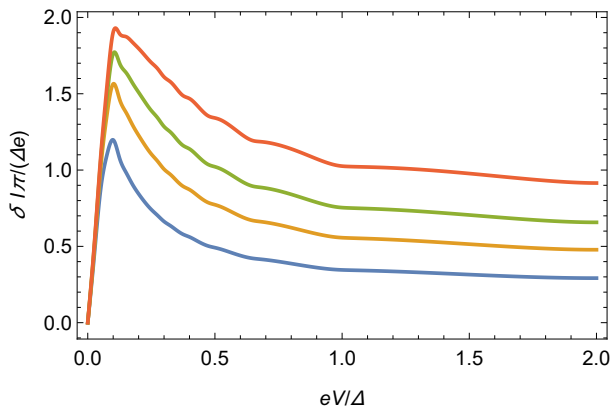}
\caption{The linear in ${\mathcal R}$ current correction $\overline{\delta I}(V)=\overline I_0(V)-\overline I(V)$ evaluated at $T=0.1 \Delta$, $\theta=0.01\Delta$ and different transmission values  ${\mathcal T}=0.97,\, 0.95,\, 0.93$ and 0.9 (bottom to top).}
\end{figure}

In addition to the above analysis we also performed a numerical calculation of $\delta I(t)$ employing the algorithm described in Appendix A.
Our numerics clearly supports the above perturbative results  and, furthermore, allows to sufficiently accurately determine the value of the prefactor $\alpha$ in Eq. (\ref{seq1}). We obtain $\alpha \simeq 3$ which is numerically close to $\alpha =\pi$. With this in mind, we may conclude that our result for $\overline{\delta I}$ is consistent with Eq. (\ref{corr}) that follows from the Landau-Zener-tunneling-type of analysis \cite{ab1}.

Our numerical results for $\overline{\delta I}$ as a function of $V$ are displayed in Fig. 3 for several different values of ${\mathcal R}$ and $eV < 2\Delta$. We observe that in accordance with Eq. (\ref{seq1}) the current correction decays as $\overline{\delta I} \propto 1/V$ roughly between $eV\approx 0.2\Delta$ and $eV\sim \Delta$. Also the subharmonic gap structure at $eV=2\Delta /k$ with $k=2,3,...$ is clearly observed, in particular for curves corresponding to lower transmissions ${\mathcal T}$. At larger voltages $eV \gtrsim 2\Delta$ (not shown in Fig. 3) the current correction $\overline{\delta I}$ starts to grow with $V$ and demonstrates the expected behavior $\overline{\delta I} \simeq e^2{\mathcal R}V/\pi$ at sufficiently large $V$.

The result of our numerical calculation for the function $F(x)$ is presented in Fig. 4 along with the analytic formula (\ref{tf}) and the function $F_{LZ}(x)$  (\ref{cab2}). For ${\mathcal R} \ll 1$ the numerical curve $F(x)$ does not depend on ${\mathcal R}$ and essentially follows the dependence (\ref{tf}). In fact, the blue curve is also described by the function of the form (\ref{tf}) provided one multiplies the last term in Eq. (\ref{tf}) by an extra numerical factor $\simeq 1.19$. Such minor difference between the two curves is, of course, of no significance and could be attributed, e.g., to neglecting Andreev reflection at overgap energies while deriving Eq. (\ref{tf}). On the other hand, both these curves differ substantially from the result of Landau-Zener-tunneling-type of analysis (\ref{cab2}).

\begin{figure}
\includegraphics[width=8cm]{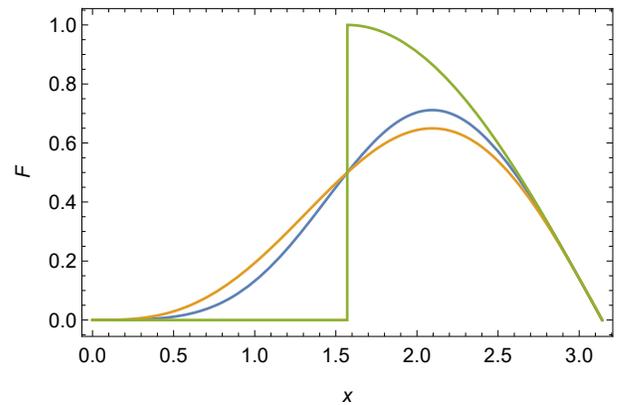}
\caption{The function $F(x)$ evaluated numerically for ${\mathcal R} \ll 1$ and $\theta=0.01\Delta$ (blue curve) and analytically (Eq. (\ref{tf}), orange curve. For comparison we also present the function $F_{LZ}(x)$ (\ref{cab2}) (green curve).}
\end{figure}

\section{Discussion}
Combining our perturbative results with those of our numerical calculation we may conclude that the perturbation theory in ${\mathcal R}$
developed here should work sufficiently well at least up to values ${\mathcal R}\lesssim 0.1$. This perturbation theory allows to microscopically derive the leading first order in ${\mathcal R}$ correction to the current. Including this correction, for ${\mathcal R} \Delta \ll e|V| \ll 2\Delta$ we have
\begin{equation}
I(t)\simeq \tilde I_c \left| \sin eVt\right|\,{\rm sgn}\,V + \frac{\alpha {\mathcal R}\Delta I_c}{2e|V|}\sin 2eVt,
\label{final}
\end{equation}
where
\begin{equation}
\tilde I_c=I_c(1-\alpha {\mathcal R}\Delta/e|V|).
\label{ren}
\end{equation}
We observe two effects. Firstly, the current $\tilde I_c$ (\ref{ren}) becomes smaller than $I_c$ since some subgap quasiparticles are eliminated from the full MAR cycle due to the presence of weak normal reflection. Secondly, these quasiparticles produce an extra Josephson-like contribution to $I(t)$ defined by the last term in Eq. (\ref{final}). Decreasing the bias voltage down to $e|V| \sim {\mathcal R}\Delta$ one reaches the limit beyond which normal reflection -- no matter how weak it is -- prevents the vast majority of subgap quasiparticles from completing the MAR cycle and the perturbation theory in ${\mathcal R}$ ceases to be valid.

Interestingly, Eq. (\ref{final}) turns out to be useful also in the latter limit of smaller $V$. Indeed, for any bias voltage $V$ the probability for quasiparticles to follow the trajectories (a) in Fig. 2 and complete the full MAR cycle is ${\mathcal T}^n$ with $n \sim \alpha \Delta/e|V|$. Accordingly, the contribution of such quasiparticles to $I(t)$ can be described by the first term in the right-hand side of Eq. (\ref{final}) also at  $e|V| < {\mathcal R}\Delta$ provided one replaces Eq. (\ref{ren}) by
\begin{equation}
\tilde I_c=I_c(1-{\mathcal R})^{\frac{\alpha \Delta}{e|V|}}.
\label{ren2}
\end{equation}
Clearly, this contribution to $I(t)$ will die out in the limit $V \to 0$ for any ${\mathcal R}>0$.

Quasiparticles following the trajectories (b) may contribute to the current only provided
the condition ${\mathcal T}^m \sim 1$ remains fulfilled, where $m$ is the total number of traverses of these quasiparticles across the weak link.
It follows immediately that for $e|V| < {\mathcal R}\Delta$ the maximum number of such traverses is $m \sim 1/{\mathcal R}$. Hence, in order to estimate the contribution of these quasiparticles to $I(t)$ it suffices to replace the number $n$ by $m$ (i.e. $\alpha \Delta/e|V| \to 1/{\mathcal R}$) in the last term of Eq. (\ref{final}). This simple estimate already yields the right order of magnitude for the current $\sim I_c$ in the limit
$e|V| \ll {\mathcal R}\Delta$. The correct current-phase dependence \cite{KO} in this regime should be recovered by additionally taking into account all higher order in ${\mathcal R}$ processes disregarded here.

Although the analysis \cite{ab1} yields the average value of the perturbative in ${\mathcal R}$ correction $\overline{\delta I}$ (\ref{corr})
which essentially matches with our Eq. (\ref{seq1}), some other features differ substantially from those found here.  Within the approach \cite{ab1} the correction to the current $\delta I(t)$ emerges due to incomplete Landau-Zener tunneling of the system to the higher Andreev level rather than due to breaking of the MAR cycle by normal reflection. As a result, the correction $\delta I(t)$ derived from this physical picture turns discontinuous (cf. Eq. (\ref{cab2})) and differs from zero only within a half of the Josephson period.

Let us evaluate the ratio between the current correction harmonics $\overline{\delta I}F_l$ and the current harmonics $I_l$ corresponding to fully open quantum point contact, Eq. (\ref{1}). With the aid of Eq. (\ref{seq2}) we get
\begin{equation}
\frac{\overline{\delta I}F_l}{I_l}\sim  \frac{{\mathcal R}\Delta}{e|V|}\left( 1+\frac{3\pi}{8}i\delta_{l,1}-\frac{3\pi}{8}i\delta_{l,-1}\right).\label{rattf}
\end{equation}
The same calculation with the function $F_{LZ}(eVt)$ (\ref{cab2}) yields
\begin{equation}
\frac{\overline{\delta I}(F_{LZ})_l}{I_l}\sim \frac{{\mathcal R}\Delta}{e|V|}\left( 1-2il(-1)^l\right). \label{ratf}
\end{equation}
Some quantitative difference between the two above expressions is observed already for the first current harmonics. For all higher harmonics with $|l| >1$ the difference between Eqs. (\ref{rattf}) and (\ref{ratf}) becomes even more pronounced, since the imaginary contribution vanishes for such $l$ in the first of these equations, while it remains non-zero and grows with $l$ in the second one.  Thus, Eq. (\ref{rattf}) assures that for ${\mathcal R}\ll 1$ and $e|V| \gg {\mathcal R} \Delta$ the perturbative correction remains small for all current harmonics. In contrast, under the same conditions the perturbation theory is obviously violated for harmonics with sufficiently large $l$ in Eq.  (\ref{ratf}).

Finally, we would also like to point out that  in many cases even equilibrium charge transport in superconducting weak links cannot be described solely in terms of Andreev bound states and their filling factors. This is the case, for instance, in ballistic SNS junctions with thicknesses of the N-layer comparable to (larger than) the superconducting coherence length or in junctions formed by two different  superconductors with $\Delta_1\gg \Delta_2$ where there exist no Andreev bound states in the range $-\pi/2<\chi <\pi/2$ at all. Accordingly, the current cannot be derived by simply taking a derivative of the Andreev states energy with respect to $\chi$. At the same time, a description of Josephson dynamics in terms of MAR is, of course, well possible also in those cases.

To conclude, it appears that one should be cautious while describing Josephson dynamics of superconducting weak links at high transmissions operating only with a pair of subgap Andreev levels and including the process of Landau-Zener tunneling between them. While this reduced physical picture is intuitively appealing and might capture some of the features, it turns insufficient for some other features, as it is demonstrated by our analysis.

\begin{widetext}
\appendix
\section{Numerical procedure}
It is convenient to rewrite a single channel version of the effective action in Eq. (\ref{ea}) in the form (cf. also \cite{Cuevas})
\begin{equation}
iS_t=\frac{1}{2}{\rm Tr}\left[ \ln \left(1+\frac{\sqrt{{\mathcal T}}}{2}\left(\check Q_L-\check Q_R \right)\right)+\ln \left(1-\frac{\sqrt{{\mathcal T}}}{2}\left(\check Q_L-\check Q_R \right)\right)\right].\label{twol}
\end{equation}
Evaluation of the current requires inverting the matrices in Eq. (\ref{twol}). This matrix inversion procedure follows closely to that of Ref. \onlinecite{Sha}, and we shall just write out the resulting expressions. The current is defined as
\begin{equation}
I(t)=\sum_{l=-\infty}^\infty I_l e^{-2ileVt}, \quad I_{l}=\frac{e^2}{\pi}{\mathcal T}V\delta_{l,0}+\int\limits_{-\infty}^\infty \frac{d\epsilon}{2\pi} \left[ {\cal I}(\epsilon,l)+{\cal I}^*(\epsilon,-l) \right],\label{symbexpr}
\end{equation}
where
\begin{eqnarray}
&& {\cal I}(\epsilon,l)=\frac{e\sqrt{{\mathcal T}}}{4}\left\{ \left(\tilde\zeta^R\left(\epsilon-\frac{eV}{2},2l\right)+
\tilde\zeta^R\left(\epsilon+\frac{eV}{2},2l\right)\right)g^K(\epsilon)\right.\label{cali}\\
&& -\left(\tilde\zeta^R
\left(\epsilon-\frac{eV}{2},2l+1\right)+
\tilde\zeta^R\left(\epsilon+\frac{eV}{2},2l-1\right)\right)f^K(\epsilon)\nonumber\\
&&+g^K(\epsilon)\sum_{m+n=2l} \left[ Y\left(\epsilon-\frac{eV}{2},n\right)\tilde\zeta^R\left(\epsilon-\frac{eV}{2},n\right)
\tilde\zeta^R\left(-\epsilon+\frac{eV}{2},m\right)\right.\nonumber\\
&& \left.-Y\left(\epsilon+\frac{eV}{2},n\right)\tilde\zeta^R\left(\epsilon+\frac{eV}{2},n\right)
\tilde\zeta^R\left(-\epsilon-\frac{eV}{2},m\right)\right]\nonumber\\
&& +f^K(\epsilon)\sum_{m+n-1=2l}
Y\left(\epsilon-\frac{eV}{2},n\right)\tilde\zeta^R\left(\epsilon-\frac{eV}{2},n\right)
\tilde\zeta^R\left(-\epsilon-\frac{eV}{2},m\right)\nonumber\\
&& \left.-f^K(\epsilon)\sum_{m+n+1=2l}
Y\left(\epsilon+\frac{eV}{2},n\right)\tilde\zeta^R\left(\epsilon+\frac{eV}{2},n\right)
\tilde\zeta^R\left(-\epsilon+\frac{eV}{2},m\right)\right\}.\nonumber
\end{eqnarray}
Both normal and anomalous Keldysh functions in Eq. (\ref{cali}) are defined by the standard relations
\begin{equation}
g^K(\epsilon)= 2\, {\rm Re} \left[ g^R(\epsilon)\right] \tanh\frac{\epsilon}{2T}, \quad f^K(\epsilon)= 2\, {\rm Re} \left[ f^R(\epsilon)\right] \tanh\frac{\epsilon}{2T}\label{kc}
\end{equation}
and the function $\tilde\zeta^R(\epsilon,n)$ takes the form
\begin{equation}
\tilde\zeta^R(\epsilon,n)=\left(-\frac{\sqrt{{\mathcal T}}}{2}\right)^{-n}\zeta^R_0(\epsilon,eV)
\prod_{k=n}^{-1}\frac{f^R\left(\epsilon+
\left(k+\frac{1}{2} \right)eV \right)}{\delta_k(\epsilon,eV)}, \quad n<0,\label{s1}
\end{equation}
\begin{equation}
\tilde\zeta^R(\epsilon,n)=\left(\frac{\sqrt{{\mathcal T}}}{2}\right)^n \zeta^R_0(\epsilon,eV)
\prod_{k=1}^{n}\frac{f^R\left(\epsilon+
\left(k-\frac{1}{2} \right)eV \right)}{d_k(\epsilon,eV)}, \quad n>0\label{s2}
\end{equation}
and
\begin{equation}
\tilde\zeta^R(\epsilon,n)=\zeta_0^R(\epsilon,eV), \quad n=0.
\end{equation}
The functions $d_k,\,\delta_k$ (cf. Ref. \onlinecite{M}) are defined by the following recurrence relations
\begin{eqnarray}
&&\delta_{-N}(\epsilon,eV)=1+\frac{\sqrt{{\mathcal T}}}{2}\left(g^R\left(\epsilon+\left(-N+\frac{1}{2}\right)eV \right)-1\right), \label{expp1}\\ &&\delta_n(\epsilon,eV)=1+\frac{\sqrt{{\mathcal T}}}{2}\left(g^R\left(\epsilon+\left(n+\frac{1}{2}\right)eV \right)-g^R\left(\epsilon+\left(n-\frac{1}{2}\right)eV \right)\right)+\nonumber\\
&& +\frac{{\mathcal T}}{4}\frac{\left[ f^R\left(\epsilon+\left(n-\frac{1}{2}\right)eV \right)\right]^2}{\delta_{n-1}(\epsilon,eV)}, \nonumber
\\
&& \delta_N(\epsilon,eV)=1+\frac{\sqrt{{\mathcal T}}}{2}\left(1-g^R\left(\epsilon+\left(N-\frac{1}{2}\right)eV \right)\right)+\frac{{\mathcal T}}{4}\frac{\left[ f^R\left(\epsilon+\left(N-\frac{1}{2}\right)eV \right)\right]^2}{\delta_{N-1}(\epsilon,eV)}.\nonumber
\end{eqnarray}
\begin{eqnarray}
&&d_{N}(\epsilon,eV)=1+\frac{\sqrt{{\mathcal T}}}{2}\left(1-g^R\left(\epsilon+\left(N-\frac{1}{2}\right)eV \right)\right),\label{exp2}\\
&&d_n(\epsilon,eV)=1+\frac{\sqrt{{\mathcal T}}}{2}\left(g^R\left(\epsilon+\left(n+\frac{1}{2}\right)eV \right)-g^R\left(\epsilon+\left(n-\frac{1}{2}\right)eV \right)\right)+\nonumber\\
&& +\frac{{\mathcal T}}{4}\frac{\left[ f^R\left(\epsilon+\left(n+\frac{1}{2}\right)eV \right)\right]^2}{d_{n+1}(\epsilon,eV)}, \nonumber\\ &&
d_{-N}(\epsilon,eV)=1+\frac{\sqrt{{\mathcal T}}}{2}\left(g^R\left(\epsilon+\left(-N+\frac{1}{2}\right)eV \right)-1\right) + \frac{{\mathcal T}}{4}\frac{\left[ f^R\left(\epsilon+\left(-N+\frac{1}{2}\right)eV \right)\right]^2}{d_{-N+1}(\epsilon,eV)}. \nonumber
\end{eqnarray}
We also have
\begin{equation}
\zeta_0^R(\epsilon,eV)=\frac{\prod_{k=1}^N d_k(\epsilon,eV)}{\prod_{k=0}^N \delta_k(\epsilon,eV)},\label{zp1}
\end{equation}
which can be rewritten as  $\zeta^R_0(\epsilon,eV)=1/X$, where
\begin{eqnarray}
X(\epsilon,eV)=d_0(\epsilon,eV)+\delta_0(\epsilon,eV)-1-\frac{\sqrt{{\mathcal T}}}{2}\left[ g^R\left(\epsilon+\frac{eV}{2} \right)-g^R\left(\epsilon-\frac{eV}{2} \right)\right].
\label{z01}
\end{eqnarray}

Finally, the function $Y(\epsilon,n)$  in Eq. \eqref{cali} is defined as
\begin{equation}
Y(\epsilon,n)=\left\{\begin{array}{l} 2d_n(\epsilon)-1+\sqrt{{\mathcal T}}g^R\left(\epsilon+\left(n-\frac{1}{2} \right)eV\right),\;{\rm if}\; n\ge 1\\
d_0(\epsilon)-\delta_0(\epsilon)+\frac{\sqrt{{\mathcal T}}}{2}\left[ g^R\left(\epsilon+\frac{eV}{2}\right) + g^R\left(\epsilon-\frac{eV}{2}\right)\right],\;{\rm if}\; n=0\\
-2\delta_n(\epsilon)+1+\sqrt{{\mathcal T}}g^R\left(\epsilon+\left(n+\frac{1}{2} \right)eV\right),\;{\rm if}\; n\le -1
\end{array} \right.\label{ydef}
\end{equation}
\end{widetext}
The functions $d_n(\epsilon)$ and $\delta_n(\epsilon)$ are also used in the definition of $\tilde\zeta^R(\epsilon,n)$ and are given by the recurrence relations  \eqref{expp1}, \eqref{exp2}.

Our numerical procedure amounts to first choosing a sufficiently large number of Andreev reflection cycles $N$ relevant in the limit of small bias voltages $eV \ll 2\Delta$. This information is included in the boundary conditions given by the first lines of Eqs. \eqref{expp1}, \eqref{exp2}.  As a next step, it is necessary to resolve the recurrence relations \eqref{expp1}, \eqref{exp2} and to construct the functions $\tilde\zeta^R(\epsilon,n)$ and  $Y(\epsilon,n)$. Then, integrating over energy in Eq. \eqref{symbexpr}, one recovers all current harmonics $I_{l}$. It is also worth pointing out that in the course of our calculation we essentially employed the standard relations between the retarded and advanced Green functions $g^A(\epsilon)=-\left(g^R(\epsilon)\right)^*,\; f^A(\epsilon)=-\left(f^R(\epsilon)\right)^*$ as well as
the conditions $g^R(-\epsilon)=(g^R(\epsilon))^*$ and  $f^R(-\epsilon)=-(f^R(\epsilon))^*$.

\end{document}